\documentclass[review]{elsarticle}

\usepackage{amssymb}

\newcommand{\clK}{{\cal K}}
\newcommand{\clD}{{\cal D}}

\newcommand{\hclL}{\hat{\cal L}}   \newcommand{\hclF}{\hat{\cal F}}

\newcommand{\rgl}{\rangle}
\newcommand{\lgl}{\langle}

\newcommand{\be}{\begin{equation}}
\newcommand{\ee}{\end{equation}}
\newcommand{\bea}{\begin{eqnarray}}
\newcommand{\eea}{\end{eqnarray}}

\begin{document}

\begin{frontmatter}

\title{Destruction of ultra-slow diffusion in a three dimensional 
cylindrical comb structure}

\author[lab1]{A. Iomin}\corref{cor}
\ead{iomin@physics.technion.ac.il}

\author[lab2]{V. M\'{e}ndez }\ead{vicenc.mendez@uab.cat}

\cortext[cor]{Corresponding author}
\address[lab1]{Department of Physics, Technion, Haifa, 32000, Israel}
\address[lab2]{Grup de F\'{\i}sica Estad\'{\i}stica, Departament
de F\'{\i}sica. Universitat Aut\`onoma de Barcelona. Edifici Cc.
08193 Cerdanyola (Bellaterra) Spain}

\begin{abstract}
We present a rigorous result on ultra-slow diffusion by solving a Fokker-Planck
equation, which describes anomalous transport in a three dimensional (3D) comb.
This 3D cylindrical comb consists of a cylinder of discs threaten on a backbone.
It is shown that the ultra-slow contaminant spreading
along the backbone is described by the
mean squared displacement (MSD) of the order of $\ln (t)$.
This phenomenon  takes place
only for normal two dimensional diffusion inside the infinite secondary
branches (discs). When the secondary branches have finite boundaries,
the ultra-slow motion is a transient process and the asymptotic behavior
is normal diffusion. In another example, when anomalous diffusion takes
place in the secondary branches, a destruction of ultra-slow
(logarithmic) diffusion takes place as well. As the result, one observes ``enhanced''
subdiffusion with the MSD $\sim t^{1-\alpha}\ln t$, where $0<\alpha<1$.

\end{abstract}

\end{frontmatter}

\section{Introduction}

A contaminant transport in inhomogeneous media exhibits anomalous diffusion.
This phenomenon is well established \cite{montrollscher,dukhne1,dukhne2,prl}
and reviewed (see \textit{e.g.}, \cite{bouchaud,isichenko,dukhne,bAH,sokolov}).
A comb model is a simple description of anomalous diffusion, which however
reflects many important  transport properties of  inhomogeneous media.
The comb model was introduced as a toy model for understanding  anomalous
transport in low dimensional percolation clusters
\cite{white-barma,weiss-havlin,baskin1}.
It is a particular example of a non-Markovian phenomenon,
which is also explained in the framework of continuous time random
walks \cite{weiss-havlin,shlesinger,klafter}.

Anomalous diffusion on the two dimensional ($2D$) comb is described
by the $2D$ probability distribution function (pdf) $P=P(x,y,t)$ of
finding a particle at time $t$ at position $y$ along the secondary
branch that crosses the backbone at point $x$. Since the transport is
inhomogeneous, the diffusion on a 2D comb-like structure is described
by the following equation
\cite{baskin1}
\begin{equation}\label{comb_eq1}
\partial_tP(x,y,t)=\mathcal{D}_{x}\delta(y)
\partial_x^{2}P(x,y,t)+\mathcal{D}_{y}
\partial_y^{2}P(x,y,t)\, .
\end{equation}
Here
$\mathcal{D}_{x}\delta(y)$ is the diffusion coefficient in the $x$ direction, and
$\mathcal{D}_{y}$ is the diffusion coefficient in the $y$ direction. The
$\delta$-function in the diffusion coefficient in the $x$ direction implies that
diffusion occurs along the $x$ direction at $y=0$ only. Thus, this
equation describes diffusion along the backbone (at $y=0$) where
the secondary branches (fingers) play a role of traps.
The comb model with infinite secondary branches (fingers)
describes subdiffusion with the mean squared
displacement (MSD), or the variance $\sigma_x^2(t)$, which spreads by power law
$\sim t^{1/2}$ \cite{white-barma,weiss-havlin,baskin1}.
For the finite secondary branches, subdiffusion is a transient process until time $T_0$,
and after a transient time scale $t>T_0$ the transport along the backbone corresponds
to normal diffusion with the MSD $\sim t$ \cite{bouchaud}.

It is convenient to work with dimensionless variables and parameters.
This can be obtained by the re-scaling with relevant combinations of the
comb parameters $[D_x]=\frac{cm^3}{sec}$ and $[D_y]=\frac{cm^2}{sec}$,
such that the dimensionless time and coordinates are
\begin{equation}\label{dimensionless}
D_x^2t/D_y^3\rightarrow t\, , ~~~D_xx/D_y\rightarrow x\, ,~~~
D_xy/D_y\rightarrow y/\sqrt{D}\, ,
\end{equation}
where $D$ can be considered
as a dimensionless diffusion coefficient for the secondary branch dynamics.
\begin{figure}[htbp]
\includegraphics[width=0.8\hsize]{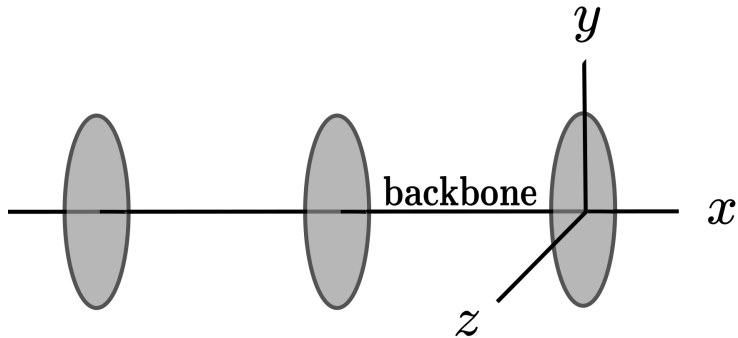}
\caption{Schematic representation of the 3D cylindrical comb.
The backbone coincides with axis $x$ and the discs belong to the $y-z$ plane.}
\label{fig:fig1}
\end{figure}

In this paper we consider anomalous ultra-slow diffusion in a three dimensional
cylindrical comb, which consists of a cylinder of discs threaded on the $x$ axis,
as it is shown in Fig.~\ref{fig:fig1}.
Recently, the ultra-slow phenomena were attracted much attention in biological search problems with long-range memories \cite{silva,boyer}.
In the continuous time random work, ultra-slow diffusion
is known as a result of super-heavy-tailed distributions of waiting times,
see details of a discussion in Ref. \cite{chechkin,denisov} and numerical
results in Ref.~\cite{bodrova}. These heavy-tailed
distributions result in the ratio $\sigma^2(t)/t^{\alpha}\rightarrow 0$ at $t\rightarrow \infty$, which tends to zero, in contrast with subdiffusion, where this ratio limits to a constant value.
We show that the MSD exhibits a logarithmic behavior in time $ \lgl x^2(t)\rgl\sim \ln(t)$, which is a result of the transversal branch dynamics in the  2D space. This behavior
has been discussed  in the framework of scaling arguments for the return probability
\cite{vulpiani,bouchaud}, which are based on the fractal dimension $d$ of the transversal
branch structure and the spectral dimension $d_s$, which is defined
by decay of the return probability $\sim t^{-d_s/2}$ \cite{bAH,alexander1982} and $\lgl x^2(t)\rgl\sim t^{1-\frac{d_s}{2}}$ for $d_s<2$ \cite{vulpiani}. As it is shown
in Ref.~\cite{vulpiani}, $\lgl x^2(t)\rgl\sim \ln t$ for $d_s=2$
Therefore, it is instructive to present a rigorous result by solving the Fokker-Planck
equation in the three dimensional space within the
cylindrical comb geometry constraint, when $d=2$.

\section{Dynamics in a cylindrical comb: an infinite comb model}\label{sec:IC}

We consider a 3D cylindrical comb \cite{dukhne,vulpiani}, shown in Fig.~\ref{fig:fig1},
in the framework of a standard formulation of the comb model
(\ref{comb_eq1}) for the 3D case.
Therefore, the random dynamics on this structure is describe by the $3D$
distribution function $P=P(x,y,z,t)$, where the $x$-axis corresponds to the
backbone, while the dynamics on the two dimensional secondary branches is described
by the $y$ and $z$ coordinates. The diffusion equation in the dimensionless
variables and parameters reads
\begin{equation}\label{kebabE}
\partial_tP=\delta(y)\delta(z)\partial_x^2P+D(\partial_y^2+\partial_z^2)P\, .
\end{equation}
The natural boundary conditions are taken at infinity,
where the distribution function and its first space derivatives vanish.
The initial condition is
\begin{equation}
P(x,y,z,t=0)=\delta(x)\delta(y)\delta(z).
\label{eq:ic}
\end{equation}

\subsection{Analysis in the time domain}\label{sec:TD}
The formal solution of Eq. (\ref{kebabE}) can be presented in a convolution form
\begin{equation}\label{convolution}
P(x,y,z,t)=\int_0^tG(y,z,t-t')F(x,t')dt'\, ,
\end{equation}
where $G(y,z,t)$ describes two dimensional diffusion in the secondary branches,
while $F(x,t)$ is a solution along the backbone. Taking into account the cylindrical
symmetry, one obtains for the $y-z$ surface
\begin{equation}\label{G}
G(y,z,t)\equiv G(r,t)=\frac{1}{4\pi Dt}\exp\Big(-\frac{r^2}{4Dt}\Big)\, ,
\end{equation}
where $r^2=y^2+z^2$.
To define the MSD in the $x$ direction, one needs to find a coarse grained distribution
$P_1(x,t)$ by integrating (\ref{convolution}) over $y$ and $z$ taking into account that the element of differential area is $dydz=d\theta rdr$. From Eq. (\ref{G}), one obtains
\begin{eqnarray}\label{P1}
P_1(x,t)&=&\int_{-\infty}^{\infty}P(x,y,z,t)dydz \nonumber \\
&=&\int_0^t
\left[\int_0^{\infty}r dr \int_0^{2\pi}d\theta G(r,t-t')\right]F(x,t')dt' \nonumber \\
&=&  \int_0^t F(x,t')dt'\, .
\end{eqnarray}
In the Laplace space, this expression establishes a relation between
$\tilde{P}_1(s)=\hclL[P_1(t)] $ and $\tilde{F}(s)=\hclL[F(t)]$. This relation reads
\begin{equation}\label{P1toF}
\tilde{F}(x,s)=s\tilde{P}_1(x,s) \, .
\end{equation}
The initial condition for the coarse grained distribution is $P_1(x,t=0)=\delta(x)$.
Using relation (\ref{P1toF}), one obtains an equation for $\tilde{P}_1(x,t)$. Integrating
Eq. (\ref{kebabE}) over $y$ and $z$, and taking into account Eqs.  (\ref{convolution}), (\ref{G}) and (\ref{P1toF}) , one obtains
\begin{equation}\label{Equation4P1}
s\tilde{P}_1=\frac{1}{4\pi D}\partial_x^2\hclL[t^{-1}]
s\tilde{P}_1(x,s)+\delta(x)\, .
\end{equation}
Note that the Laplace transform of $t^{-1}$ exists as a principal
value integral \cite{bateman-f}.
Fourier transforming Eq. (\ref{Equation4P1}), one obtains
\begin{equation}\label{FLP1}
\bar{\tilde{P}}_1(k,s)=\frac{4\pi D}{s(4\pi D+k^2\hclL[t^{-1}])}\, .
\end{equation}
This yields the MSD in the form
\begin{equation}\label{MSD}
\langle x^2(t)\rangle
= \hclL^{-1}\Big[-\frac{d^2}{d\,k^2}\bar{\tilde{P}}_1(k,s)\Big]_{k=0}
=\frac{1}{2\pi D}\int_{-i\infty}^{+i\infty}\hclL[t^{-1}]\frac{e^{st}ds}{s}\, .
\end{equation}
Taking into account that $\hclL^{-1}\hclL\Big[t^{-1}\Big]\equiv t^{-1}$ and $e^{st}/s=\int^te^{st}dt +C$,
where $C$ is an unnecessary/unimportant constant of the indefinite integration, one obtains
\begin{eqnarray}\label{MSD-final}
\langle x^2(t)\rangle&=&\frac{1}{2\pi D} \ln(t) +\frac{C}{2\pi D\,t}\nonumber\\
&=&\frac{1}{2\pi D} \ln(t)\, , \quad \mbox{as} \quad t\rightarrow \infty\, .
\end{eqnarray}
Therefore, for the large time dynamics, ultra-slow diffusion
takes place with the MSD being of the order of $\sim \ln(t)$.

\subsection{Consideration in the Laplace domain}\label{sec:LD}

Let us consider a relation between the temporal dynamics and the dynamics in the
Laplace space.
Performing the Laplace transform of Eq. (\ref{kebabE}), one obtains
\begin{equation}\label{LDomainComb}
s\tilde{P}=\delta(y)\delta(z)\partial_x^2\tilde{P}+D(\partial_y^2+\partial_z^2)\tilde{P}+P_0\, .
\end{equation}
Correspondingly, Eq. (\ref{convolution}) reads in the Laplace domain
\begin{equation}\label{tildaGF}
\tilde{P}(x,y,z,s)=\tilde{G}(y,z,s)\tilde{F}(x,s)\equiv\tilde{G}(r,s)\tilde{F}(x,s) \, ,
\end{equation}
where the cylindrical symmetry is taken into account in the last term.
Taking into account Eqs. (\ref{LDomainComb}) and (\ref{tildaGF}),
one finds the solution for $\tilde{G}(r,s)$ from the equation
\begin{equation}\label{BesselEq}
u^2\tilde{G}^{\prime\prime}+u\tilde{G}^{\prime}-u^2\tilde{G}=0\, ,
\end{equation}
where prime means a derivative over $u$. This is an equation for the modified Bessel
functions $ I_0(u)$ and $K_0(u)$, where $u=r \sqrt{s/D}$  (see \textit{e.g.}, \cite{yel}).
The solution, which satisfied the boundary condition at infinity $r=\infty$, is the
modified Bessel function of the second kind
\begin{equation}\label{BesselSol}
\tilde{G}(r,s)=A\cdot K_0\Big(r\sqrt{s/D}\Big)\, .
\end{equation}
It should be stressed that the Laplace inversion of $K_0\Big(r\sqrt{s/D}\Big)$ is exactly
the solution $G$ in Eq. (\ref{G}):
\[
\int_0^{\infty}\frac{1}{4\pi D t}\exp\Big(-\frac{r^2}{4Dt}-st\Big)dt=
\frac{1}{2\pi D}K_0\Big(r\sqrt{s/D}\Big)\, .
\]
Therefore, $A=1/2\pi D$ that satisfies the normalization condition and the initial condition
for $G(y,z,t)$.
Taking into account solution (\ref{BesselSol}), we establish the relation
(\ref{P1toF}) by integrating Eq. (\ref{tildaGF}) over $y$ and $z$. Using a
property of integration of the modified Bessel function:
\begin{equation}\label{BI}
\int_0^{\infty}uK_0(au)du =1/a^2\, ,
\end{equation}
one obtains
\begin{equation}\label{tildaP1_tildaF}
\tilde{P}_1(x,s)=\tilde{F}(x,s)\int dydz\tilde{G}(x,y,s)
 = \frac{\tilde{F}}{D}\int_0^{\infty}K_0\Big(r\sqrt{s/D}\Big)rdr=\frac{\tilde{F}(x,s)}{s} \, ,
\end{equation}
which coincides exactly with the result in Eq. (\ref{P1toF}), and where we also use the Laplace transform of
$\tilde{P}_1(x,s)=
2\pi\tilde{F}\int_0^\infty\tilde{G}(r,s)rdr$ in Eq. (\ref{P1}).

Now we admit an important point of the analysis: \textit{namely Eq.
(\ref{LDomainComb}) cannot be
integrated over the $y$ and $z$, because $\tilde{G}(r,s)$ does not exist at $r=0$ (it is singular).
Therefore, to continue the analysis, one has to return to the time domain and repeat
the analysis for the temporal dynamics, performed in section \ref{sec:TD}}.
This situation differs cardinally from the analysis for the $2D$ comb (\ref{comb_eq1}),
where the finite expression for MSD can be obtained in the Fourier-Laplace domain.

\section{Comb dynamics with finite discs: Transition to normal diffusion}

To consider anomalous diffusion on finite combs, we consider
reflected boundary conditions at $r=R$, such that $\partial_r\tilde{G}(r=R,s)=0$,
which determines the absence of the probability flux in the direction
normal to the boundary surface.
In this case, solution of Eq. (\ref{BesselEq}) is found in the form of modified Bessel
function of the first kind $I_0(u)$. Namely, it reads
\begin{equation}\label{finite-Bessel-Sol}
\tilde{G}(r,s)=I_0\Big[(R-r)\sqrt{s/D}\big]/I_0\Big(R\sqrt{s/D}\Big)\, ,
\end{equation}
which satisfies the boundary condition
\begin{equation}\label{finite-boundary}
\frac{d}{d\,r}I_0\Big[(R-r)\sqrt{s/D}\big]\Big|_{r=R}=
I_1\Big[(R-r)\sqrt{s/D}\big]|_{r=R}=0\, ,
\end{equation}
while for $r=0$, one obtains $\tilde{G}(0,s)=1$. This corresponds to a standard
construction of the solution \cite{baskin1,bi2004}.

Integration of Eq. (\ref{LDomainComb}) over the $y$ and $z$ yields
\begin{equation}\label{finite-LDomainComb}
s\tilde{P}_1(x,s)=\partial_x^2\tilde{F}(x,s)+1\, .
\end{equation}
Again, the relation between $\tilde{P}_1$ and $\tilde{F}$
can be established by integration of  $I_0(u)$ over the $y-z$ surface of discs,
which yields
\[2\pi\int_0^{R}r I_0\Big[(R-r)\sqrt{s/D}\Big]dr=2\pi R^2\int_0^1uI_0[a(1-u)]du\, .\]
Here we used the following variable changes $u=r/R$ and $a=R\sqrt{s/D}$.
Then using another variable change $w=1-u$, one obtains two table integrals \cite{gradshteyn}
\[
\frac{R^2}{a}\int_0^aI_0(w)dw - R^2\int_0^1wI_0(aw)dw
=\frac{R^2}{a}
\Big[2\sum_{n=0}^{\infty}I_{2n+1}(a) - I_1(a)\Big]\, .
\]
Here, we are interested in the long time dynamics, when $s\rightarrow 0$ and $a\ll 1$, correspondingly.
In this case $I_n(a)\approx \Big(\frac{a}{2}\Big)^n/\Gamma(n+1)$, and we take into account
the first term with $n=0$ in the sum that yields $I_1(a)$ in the squared brackets.
Finally, one obtains for the long time asymptotics\footnote{This result can be obtained
from Eq. (\ref{finite-Bessel-Sol}) taking into account that $\tilde{G}(r,s)\approx 1$
for the small argument in the limit $s\rightarrow 0$.}
\begin{equation}\label{finite-integration}
\int_0^R\tilde{G}(r,s)dr\approx \pi R^2\, .
\end{equation}
This yields the following relation
\begin{equation}\label{finite-P1toF}
\tilde{F}(x,s)=\frac{\pi}{R^2}\tilde{P}_1(x,s)\, .
\end{equation}
Substituting relation (\ref{finite-P1toF}) in Eq. (\ref{finite-LDomainComb})
and performing the
Laplace inversion, one obtains the Fokker-Planck equation for normal diffusion
with the diffusion coefficient $\pi/R^2$
\begin{equation}\label{FPE-normalDif}
\partial_tP_1=\frac{\pi}{R^2}\partial_x^2P_1\, .
\end{equation}
It is worth stressing that this long time diffusion takes place  only for times
larger then a transient time $t>t_0$, where $t_0=R^2/D$. This situation is different
from the long-time asymptotics observed in \cite{chechkin}.

This result is generic for combs with finite secondary branches
(either fingers in the 2D comb, or discs in the 3D comb).
However, the finite boundary conditions for the $y-z$ discs result in the
destruction of ultra-slow diffusion in the $x$ direction, as well.
Mathematically, this fact follows immediately from the Laplace
image $\tilde{G}(r,s)$, which depends on the boundary conditions.
The ultra-slow motion takes place only for normal diffusion in the infinite discs.

Note that diffusion in the side branch discs can be anomalous as well.
Does this ultra-slow diffusion survives in this case? The answer is it does not.
We prove this statement in the next section.

\section{Anomalous diffusion in discs}\label{sec:ADD}

What happens with this ultra-slow diffusion if
diffusion in the transversal disks is anomalous and described by a
memory kernel $\clK(t)$? The comb model (\ref{kebabE}) now reads
\begin{equation}\label{ADD-kebabE}
\partial_tP=\delta(y)\delta(z)\partial_x^2P+D(\partial_y^2+\partial_z^2)
\int_0^t\clK(t-t')P(t')dt'\, .
\end{equation}
The temporal kernel $\clK(t)$ is defined in the Laplace domain through a
waiting time pdf $\psi(t)$ \cite{klafter,MFH}
\begin{equation}\label{ADD-K}
\tilde{\clK}=s\tilde{\psi}(s)/[1-\tilde{\psi}(s)]\, .
\end{equation}
Repeating the analysis in the Laplace domain of Sec.~\ref{sec:LD}, one obtains solution
(\ref{BesselSol}) in the form
\begin{equation}\label{ADD-BesselSol}
\tilde{G}(r,s)=A\cdot K_0(Br)\, ,
\end{equation}
where $B=\sqrt{s/D\clK(s)}$ and $A$ is a normalization constant.
Therefore integration (\ref{BI}) yields
\begin{equation}\label{ADD-P1toF}
\tilde{F}(x,s)=\frac{B^2}{A}\tilde{P}_1(x,s)\, .
\end{equation}

As already admitted above (in Sec.~\ref{sec:LD}), $\tilde{G}(r,s)$ is
singular function at $r=0$. Therefore, as in Eq. (\ref{LDomainComb}),
straightforward integration of Eq. (\ref{ADD-kebabE}) over the $y$ and $z$
coordinates can be performed only in the real time domain.
However, the function $G(r=0,t)$ does exists and correspondingly
$\hclL[G(r=0,t)](s)$ exists as well, at least as a principal value
integral like in Eq. (\ref{Equation4P1}). Therefore,
to obtain equation for $P_1$, one has to return to the time domain
consideration for $P_1(r,t)$ by integrating Eq. (\ref{ADD-kebabE})
over $y$ and $z$. To be specific, let us consider subdiffusion in the $y-z$ discs,
described by  $\tilde{\psi}(s)=\frac{1}{1+(\tau s)^{\alpha}}$
and correspondingly with the memory kernel
$$\tilde{\clK}(s)=s^{1-\alpha}/\tau^{\alpha}\, ,$$
where $\tau$ is a dimensionless  characteristic time scale and $0<\alpha<1$.
In this case $B^2=s^{\alpha}\tau^{\alpha}/D$.
Therefore, in the limit $r\rightarrow 0$ the argument $Br\ll 1$ and
Eq. (\ref{ADD-BesselSol}) reads for this small argument \cite{yel}
\begin{equation}\label{ADD-BesselSmall}
K_0[(bs)^{\alpha/2}]\approx\ln\frac{2}{\gamma}-\frac{\alpha}{2}\ln(bs)\, ,
\end{equation}
where $\gamma$ is an Euler constant and $b=\tau (r^2/D)^{1/\alpha}$.
Now, we perform the Laplace inversion at $r\rightarrow 0$
\begin{equation}\label{ADD-LI}
G(r,t)= A\int_{-i\infty}^{i\infty}K_0[(bs)^{\alpha/2}]e^{st}ds
\approx A\delta(t)\ln\frac{2}{\gamma}
- Ab^{-1}\frac{\alpha}{2}\int_{-i\infty}^{i\infty}\ln (p) e^{pt_b} dp\, ,
\end{equation}
where $t_b=t/b$. The last term can be presented in a form of a table integral
\cite{bateman-f}
$$\int_{-i\infty}^{i\infty}\ln (p) e^{pt_b} dp=
\frac{d}{d\,t_b}\int_{-i\infty}^{i\infty}p^{-1}\ln (p) e^{pt_b} dp=-\frac{1}{t_b}\,. $$
Finally, one obtains for $t>0$ and $r\rightarrow 0$
\begin{equation}\label{ADD-Gapprox}
G(r\rightarrow 0,t)\approx \frac{A\alpha}{2}\cdot \frac{1}{t}\, .
\end{equation}
This result is independent of $r$ and  therefore, the limit $r=0$ is correct. For this
approximate solution, the constant $A=\frac{1}{2\pi D}$ is taken to satisfy
the limit $\alpha=1$, which corresponds to solution (\ref{G}) at $r=0$.

Repeating procedures of Sec.~\ref{sec:IC}, namely performing first integration
over $y$ and $z$ in Eq. (\ref{ADD-kebabE}) and then
the Laplace transform over time, and Fourier transform over $x$, and taking into account the result of Eq. (\ref{ADD-P1toF}), one obtains a modification of Eq. (\ref{FLP1}). This reads
for $\bar{\tilde{P}}_1(k,s)=\hclL\hclF[P_1(x,t)]$
\begin{equation}\label{ADD-FLP1}
\bar{\tilde{P}}_1(k,s)=\frac{1}{s+\bar{\clD}_{\alpha}s^{\alpha}k^2\hclL[t^{-1}])}\, ,
\end{equation}
where $\bar{\clD}_{\alpha}=\alpha\tau^{\alpha}/D$.
Repeating the argument for the inferring Eq. (\ref{MSD-final}), one obtains
for the MSD
\begin{equation}\label{ADD-MSD}
\langle x^2(t)\rangle
=\frac{2\bar{\clD}_{\alpha}}{\Gamma(1-\alpha)}\int_0^t\frac{\ln(t')dt'}{(t-t')^{\alpha}}
\propto \clD_{\alpha}t^{1-\alpha}\ln(t)\, .
\end{equation}
Note that we omit here a term $\sim\gamma t^{1-\alpha}$, which is a slower
contribution to anomalous diffusion than the term accounted in Eq. (\ref{ADD-MSD}).
Here $\clD_{\alpha}=2\bar{\clD}_{\alpha}/\Gamma(2-\alpha)$ is a
generalized transport coefficient.
Subdiffusion with the transport exponent $1-\alpha$ is dominant.
Therefore, we conclude that ultra-slow
diffusion $\sim \ln(t)$ takes place only for $\alpha=1$ that can be
realized as the result of
normal diffusion in the infinite secondary branched discs.

\section{Conclusion}
We present a rigorous result on ultra-slow diffusion by solving
the Fokker-Planck equation in the 3D cylindrical comb geometry.
It is shown that the ultra-slow motion with the MSD on the $x$ backbone
is of the order of $\ln (t)$, and it results from normal diffusion
in the secondary branched discs of the infinite radius. If the
transport in the secondary branches is anomalously diffusive (subdiffusive),
the anomalous transport becomes dominant in the backbone, as well.
As the result, ultra-slow diffusion is replaced by the anomalous
transport with the MSD $\sim t^{1-\alpha}\ln (t)$, which is more
sophisticated than usual power law subdiffusion, and we call it
\textit{enhanced subdiffusion}. This continuous transition from
the ultra-slow motion for $\alpha =1$ to enhanced subdiffusion
with $0<\alpha\leq 1$ is due to anomalous diffusion in
the secondary branched dynamics, which is controlled by the transport
exponent $\alpha$. This solution can be helpful
to model ecological processes like in generalization of an elephant random walk model \cite{silva}.

Another important modification of the model is a choice of the boundary
conditions at finite radius of the discs, which is a realistic situation.
In this case, the physical realization of the ultra-slow transport is restricted
by the transient time scale $t<t_0=\pi R^2/D$.
The important and technically specific point of the analysis is the singularity of $\tilde{G}(r,s)$ at $r=0$. The integration of the comb equations (\ref{kebabE}) and (\ref{ADD-kebabE}) over the $y$ and $z$  coordinates is performed in the time domain,
while the relation between the coarse grained pdf $P_1(x,t)$ and the backbone pdf
$P(x,r=0,t)$ is established in the Laplace domain.

A.I. thanks the Universitat Aut\`onoma
de Barcelona for hospitality and financial support,
as well as the support by the
Israel Science Foundation (ISF-1028). V.M. has been supported by the Ministerio de Ciencia e Innovaci\'on under Grant No. FIS2012-32334. V.M. also thanks the Isaac Newton Institute for Mathematical Sciences, Cambridge, for support and hospitality during the CGP programme where part of this work was undertaken.

\end{document}